\documentclass[preprint,preprintnumbers,superscriptaddress,amsmath,amssymb]{revtex4}
\usepackage{graphicx}% Include figure files
\usepackage{bm}% bold math

\begin{document}

\title{Dynamics of Quantum Fisher information in a spin-boson model}

\author{Xiang Hao}
\email{110523007@suda.edu.cn}
\affiliation{Department of Physics, Renmin University of China, Beijing 100872, China}
\affiliation{School of Mathematics and
Physics, Suzhou University of Science and Technology, Suzhou,
Jiangsu 215011, People's Republic of China}

\author{Ning-Hua Tong}
\affiliation{Department of Physics, Renmin University of China, Beijing 100872, China}

\author{Shiqun Zhu}
\affiliation{School of Physical Science and Technology, Soochow
University, Suzhou, Jiangsu 215006, People's Republic of China}

\begin{abstract}

Quantum Fisher information characterizes the phase sensitivity of qubits in the spin-boson model with a finite bandwidth spectrum. In contrast with Markovian reservoirs, the quantum Fisher information will flow from the environments to qubits after some times if the bath parameter $s$ is larger than a critical value which is related to temperatures. The sudden change behavior will happen during the evolution of the quantum Fisher information of the maximal entanglement state in the non-Markovian environments. The sudden change times can be varied with the change of the bath parameter $s$. For a very large number of entangled qubits, the sudden change behavior of the maximal quantum Fisher information can be used to characterize the existence of the entanglement. The metrology strategy based on the quantum correlated state leads to a lower phase uncertainty when compared to the uncorrelated product state.

\end{abstract}

\maketitle

\section{Introduction}

Fisher information is a key quantity for extracting the information about a parameter from a measurement-induced probability distribution ~\cite{Helstrom76,Holevo82,Giovannetti11}. In the classical realm, the Fisher information can provide a basic lower bound of the variance of any unbiased estimator due to the Cra\'{m}er-Rao inequality \cite{Braunstein94}. A large value of Fisher information represents an attainable measurement with a high precision. The quantum Fisher information is referred to as the natural extension of the classic Fisher information. Among many versions of quantum Fisher information, there is a famous definition based on the symmetric logarithmic derivatives \cite{Helstrom76,Holevo82}. In a practical estimation, the quantum Fisher information can be used to describe the sensitivity of a quantum state with respect to a parameter such as the frequency of the system \cite{Huelga97,Chin12}, the strength of the changing external field \cite{Invernizzi08,Sun10}, and the speed of the quantum evolution \cite{Frowis12}.

To improve the precision of parameter estimation, some research groups have put forward entanglement-enhanced metrological schemes \cite{Mubner92,Bollinger96, Dowling98,Brukner06,Cho06,Jin07,Esteve08,Pezze09}. It is proved that the scaling with the number of entangled particles $N$ can achieve the Heisenberg limit, proportional to $1/N$, which overcomes the shot-noise limit (or called as standard quantum limit), proportional to $1/\sqrt{N}$. However, the practical quantum technology unavoidably induces the decoherence and dissipation owing to the coupling of the system to the environment \cite{Weiss99,Breuer01}. So far, much attention has been focused on quantum metrology in the presence of Markovian or non-Markovian noise \cite{Huelga97,Chin12,Ma11,UlamOrgikh01,Shaji07,Monras07,Li08,DemkowiczDobrzanski09,Lee09,Watanabe10,Hyllus10,Kacprowicz10,Escher11}. In some Markovian channels, there is a sudden change of the quantum Fisher information \cite{Ma11}. A local dephasing environment was also taken into account in order to exactly obtain the scaling law of the parameter estimation \cite{Chin12}. It was found out that the non-Markovian dephasing environment enables quantum entangled states to outperform the strategy based on product states. In other realistic environments, we need to study if this correlated-state metrology strategy is superior to the strategy based on product states. Besides it, we want to investigate if the sudden change behavior of the quantum Fisher information happen in non-Markovian environments.

The paper is organized as follows. We introduce the definition of quantum Fisher information in Sec. II. In Sec. III, a general non-Markovian depolarizing channel is presented by a qubit coupled to a heat bath. In, Sec. IV, the crossover between non-Markovian dynamics and Markovian ones will be manifested by the quantum Fisher information with respect to a phase parameter. The sensitivity of the phase estimation based on the $N$-particle maximally entangled state is also studied in contrast with the product state. Finally, a simple discussion is concluded in Sec. V.

\section{quantum Fisher information}

In metrology, one of basic measurements is the phase estimation with respect to a unitary transformation of a linear interferometer. The phase $\phi$ is related to an unbiased estimator $\hat{\phi}$, i.e., $\langle \hat{\phi} \rangle=\phi$. The measured state is obtained by $\rho_{\phi}=e^{i\phi \hat{J}_{\vec{n}}}\rho e^{-i\phi \hat{J}_{\vec{n}}}$ where $\rho$ is the input state. $\hat{J}_{\vec{n}}$ is a generator which describes the angular momentum operator along a direction $\vec{n}=(n_1,n_2,n_3)$. According to the quantum Cra\'{m}er-Rao theorem \cite{Helstrom76,Holevo82}, the precision of the phase parameter $\phi$ has a lower bound limit, which is determined by
\begin{equation}
\label{eq:1}
\Delta \hat{\phi}\geq \Delta \phi_{QCR}=\frac {1}{\sqrt{\nu F(\rho, \hat{J}_{\vec{n}})}},
\end{equation}
where $\nu$ is the number of the experiments. And $F(\rho, \hat{J}_{\vec{n}})$ denotes the quantum Fisher information, which can be defined as
\begin{equation}
\label{eq:2}
F(\rho, \hat{J}_{\vec{n}})=\mathrm{Tr}(\rho_{\phi}L^2_{\phi}).
\end{equation}
$L_{\phi}$ is the symmetric logarithmic derivative with respect to the phase parameter $\phi$ and satisfies that $\frac {\partial \rho_{\phi}}{\partial \phi}=\frac 12(\rho_{\phi}L_{\phi}+L_{\phi}\rho_{\phi})$. To emphasize the effects of the quantum Fisher information, we set $\nu=1$ in the paper. Using the above equations, we calculate the quantum Fisher information as
\begin{equation}
\label{eq:3}
F(\rho, \hat{J}_{\vec{n}})=2\sum_{i\neq j}\frac {(\lambda_i-\lambda_j)^2}{\lambda_i+\lambda_j}\vert \langle i| \hat{J}_{\vec{n}}|j \rangle \vert^2=\vec{n}\hat{C}\vec{n}^{\dag},
\end{equation}
where $\vert i\rangle$ is the eigenvector of the density matrix $\rho$ with the corresponding eigenvalue $\lambda_i$, and the elements of the symmetric matrix $\hat{C}$ are expressed as $C_{kl}=\sum_{i\neq j}\frac {(\lambda_i-\lambda_j)^2}{\lambda_i+\lambda_j}(\langle i| \hat{J}_{k}|j \rangle \langle j| \hat{J}_{l} | i\rangle+\langle i| \hat{J}_{l}|j \rangle \langle j| \hat{J}_{k} | i\rangle)$.

For a pure state, $F(\rho, \hat{J}_{\vec{n}})=4(\Delta \hat{J}_{\vec{n}})^2$. To maximize the quantum Fisher information, we need to select the rotation along an optimal direction $\vec{n}^{o}$ by diagonalizing the symmetric matrix as $\hat{C}^{d}=\hat{O}^{\dag}\hat{C}\hat{O}=\textrm{diag}(C_1,C_2,C_3)$. $\hat{O}$ is a transformation matrix which is composed of three orthogonal and normalization eigenvectors of $\hat{C}$. According to the result of \cite{Ma11}, the maximum of the quantum Fisher information is written as
\begin{equation}
\label{eq:5}
F_{\mathrm{max}}=4\mathrm{max}(C_1,C_2,C_3).
\end{equation}
The optimal direction $\vec{n}^{o}$ is determined by the eigenvector of $\hat{C}$ with the maximal eigenvalue.

For the total state $\rho^{N}$ of $N$ independent qubits, the maximal mean value of the quantum Fisher information can be introduced as
\begin{equation}
\label{eq:6}
\bar{F}_{\mathrm{max}}=\frac {F_{\mathrm{max}}}{N}.
\end{equation}
Here, the unitary transformation related to the phase parameter $\phi$ is presented as $\exp(i\phi\sum_{i=1}^{N}\hat{J}^{i}_{\vec{n}^{o}})$ where $\hat{J}^{i}_{\vec{n}^{o}}$ is the angular momentum operator for the $\mathrm{ith}$ qubit along the optimal direction $\vec{n}^{o}$. When the values $\bar{F}_{\mathrm{max}}>1$, we have $\Delta \phi_{\mathrm{QCR}}<1/\sqrt{N}$. In this condition, the ultimate estimation limit is superior to the standard quantum limit. Only if $\bar{F}_{\mathrm{max}}\simeq N$, the Heisenberg limit of $\Delta \phi_{\mathrm{QCR}}\simeq1/N$ can be attained.

\section{non-Markovian spin-boson model}

A spin-boson system is a simple model that describes a effective two-level system coupled to a bosonic reservoir \cite{Breuer01,Wang13}. The model has extensively been applied to noisy quantum dots \cite{Hur04}, decoherence of qubits in quantum computation \cite{Thorwart01,Costi03}, quantum impurities and charge transfer in donor-acceptor systems \cite{Tornow06}. The dissipative environment in the spin-boson model can be characterized by the structured spectral function $J(\omega)$ with frequency behavior of $J(\omega)\propto\omega^{s}\omega_c^{1-s}\exp(-\omega/\omega_c)$. $\omega_c$ is the cutoff frequency of the spectrum. With the change of the bath parameter $s$, the environments vary from sub-Ohmic ones $(s<1)$ to Ohmic $(s=1)$ and super-Ohmic ones $(s>1)$. We will employ the spin-boson model to construct a general depolarizing channel.

The Hamiltonian of the spin-boson model can be described as
\begin{equation}
\label{eq:7}
H=H_{0}+H_{E}+H_{I},
\end{equation}
where $H_{0}=\frac{\omega_0}{2}\hat{\sigma}_x $ is a local hamiltonian of the system with the tunneling frequency of $\omega_0$, and $H_{E}=\sum_{k}\hat{b}_{k}^{\dag}\hat{b}_{k}$ denotes the hamiltonian of the environment including all degrees of freedom. The interaction hamiltonian between the spin and environment is expressed as $H_{I}=\sum_{k}g_{k}(\hat{\sigma}_{+}\hat{b}_{k}+\hat{\sigma}_{-}\hat{b}_{k}^{\dag})$, where the strength of the couplings can be described by the spectral function as $\sum |g_{k}|^2\rightarrow \int J(\omega)\delta(\omega-\omega_{k})d\omega$. $\hat{\sigma}_{\pm}$ are the rising and lowering operators. For the $\mathrm{kth}$ mode field, $\hat{b}_{k}$ and $\hat{b}_{k}^{\dag}$ represent the annihilation and creation operator respectively. Using a transformation $\hat{U}$, we can diagonalize the system hamiltonian into $\bar{H}_{0}^{eff}=\hat{U}^{\dag}H_{0}\hat{U}=\frac {\omega_0}{2}\hat{\bar{\sigma}}_z$. The spin operators are also mapped into the new ones, i.e., $\hat{\bar{\sigma}}_j=\hat{U}^{\dag}\hat{\sigma}_j\hat{U},~(j=z,\pm)$. We can obtain the time-convolutionless master equation of the density matrix of the spin in the interaction picture as
\begin{equation}
\label{eq:8}
\frac {d \rho(t)}{dt}=-i[\bar{H}_{0}^{eff}, \rho(t)]+\hat{L}[ \rho(t)],
\end{equation}
where $\rho(t)$ is the density matrix of the system after the unitary transformation. The Lindblad superoperator $\hat{L}$ can be written as
\begin{equation}
\label{eq:9}
\hat{L}[\rho(t)]=\sum_{m=z,\pm}\gamma_{m}(t)[\hat{\bar{\sigma}}_{m} \rho(t) \hat{\bar{\sigma}}^{\dag}_{m}-\frac 12\{\hat{\bar{\sigma}}^{\dag}_{m}\hat{\bar{\sigma}}_{m},  \rho(t) \}].
\end{equation}
We need notice that the master equation is obtained in the secular approximation \cite{Breuer01} where the high-frequency oscillating terms are neglected. The secular approximation is reasonable under the assumption of the weak coupling between the system and the environment. The time-dependent decaying parameters at a finite temperature $T$ are expressed as
\begin{widetext}
\begin{eqnarray}
\label{eq:10}
\gamma_{z}(t,T)&=&\frac 12 \int J(\omega)\coth(\frac {\hbar\omega}{2\kappa_{B}T})\frac {\sin\omega t}{\omega} d\omega \nonumber \\
\gamma_{\pm}(t,T)&=&\frac 12\int J(\omega)[(n_{T}+1)\frac {\sin(\omega \pm \omega_{0}) t}{\omega \pm \omega_{0}}+n_{T}\frac {\sin(\omega \mp \omega_{0}) t}{\omega \mp \omega_{0}}]d \omega.
\end{eqnarray}
\end{widetext}
The mean number of the field is $n_{T}=[\exp(\hbar\omega/\kappa_{B}T)-1]^{-1}$. The analytical expression of $\rho(t)$ can be written in the Hilbert space spanned by $\{ |1\rangle, |0\rangle \}$ as
\begin{equation}
\label{eq:11}
\left(
\begin{array}{cc}
(a-b)\rho_{11}(0)+b~& c\rho_{10}(0)\\
c^{\ast}\rho_{01}(0)~& (1-a)+(a-b)\rho_{00}(0)
\end{array} \right),
\end{equation}
where $\rho_{ij}(0)$ is the element of the density matrix $\rho(0)$ at the initial time and $|1(0)\rangle$ denotes the eigenvector of the new operator $\hat{\bar{\sigma}}_{z}$ with the corresponding eigenvalue $\pm 1$. The decoherence parameters are obtained as
\begin{eqnarray}
\label{eq:12}
a&=&\frac 12[(1+\delta e^{-\Gamma})+e^{-\Gamma}]\nonumber \\
b&=&\frac 12[(1+\delta e^{-\Gamma})-e^{-\Gamma}]\nonumber \\
c&=& e^{-\Lambda-i\omega_0 t},
\end{eqnarray}
where $\Gamma(t)=\int_0^{t}[\gamma_{+}(t^{\prime})+\gamma_{-}(t^{\prime})]d t^{\prime}$, $\Lambda(t)=\frac 12 \Gamma(t)+2\int_{0}^{t}\gamma_{z}(t^{\prime})d t^{\prime}$ and $\delta(t)=\int_{0}^{t}e^{\Gamma(t^{\prime})}[\gamma_{+}(t^{\prime})-\gamma_{-}(t^{\prime})]d t^{\prime}$. If $a+b=1$ and $a-b=c$, the dynamical map will be reduced to a depolarizing channel where an initial state $\rho(0)$ will evolve into a mixture of $\rho(0)$ and maximally mixed state $\hat{I}$.

\section{dynamical crossover and phase estimation}

In this general depolarizing environments, the quantum Fisher information for time-dependent mixed state $\rho(t)$ can be analytically obtained as follows. We may use the Bloch vector $\vec{B}(t)$ to simplify the description of the density matrix of single qubit as
\begin{equation}
\label{eq:13}
\rho(t)=\frac {\hat{I}+\hat{\vec{\sigma}}\cdot \vec{B}(t)}{2},
\end{equation}
where the three components of the Bloch vector is written as
\begin{eqnarray}
\label{eq:14}
B_{1}(t)&=&e^{-\Lambda}\sin \theta \cos (\omega_0 t+\varphi),\nonumber \\
B_{2}(t)&=&e^{-\Lambda}\sin \theta \sin (\omega_0 t+\varphi),\nonumber \\
B_{3}(t)&=&e^{-\Gamma}(\cos \theta +\varphi).
\end{eqnarray}
Here, the Bloch vector can also be expressed as $\vec{B}(t)=\vert \vec{B}(t) \vert(\sin \alpha \cos \beta,\sin \alpha \sin \beta,\cos \alpha)$. These time-dependent angle parameters satisfy the initial conditions of $\alpha(0)=\theta,~\theta \in [0,\pi]$ and $\beta(0)=\varphi,~\varphi \in [0,2\pi]$ . The maximal quantum Fisher information is obtained as
\begin{equation}
\label{eq:15}
F_{\mathrm{max}}(t)=e^{-2\Lambda}\sin^{2} \theta +e^{-2\Gamma}(\cos \theta+\delta)^2.
\end{equation}
To obtain the maximal quantum Fisher information, we can apply the two different kinds of the optimal rotations where the directions are expressed as
\begin{eqnarray}
\label{eq:16}
\vec{n}^{o}_{\parallel}&=&(\sin \beta, \cos \beta, 0),\nonumber \\
\vec{n}^{o}_{\perp}&=&(0,0,1).
\end{eqnarray}
In regard to all possible initial states, we use the average value of $F_{\mathrm{max}}$ which is defined as
\begin{equation}
\label{eq:17}
F^{\mathrm{A}}_{\mathrm{max}}=\frac {1}{4\pi}\int_{0}^{2\pi}\int_{0}^{\pi}F_{\mathrm{max}}\sin \theta d \theta d\varphi.
\end{equation}

Actually, the non-monotonic behaviors of the quantum Fisher information can be used to define and quantify the non-Markovianity of quantum dynamics \cite{Lu10}. The flow of the maximal quantum Fisher information for a qubit at a finite temperature is shown in Fig. 1(a). From Fig. 1(a), we can see that the large negative values of $\frac {\partial F^{\mathrm{A}}_{\mathrm{max}}}{\partial t}$ occur if the bath parameter is less than a temperature-dependent critical value, i.e., $s<s_{c}$. This also means that the quantum Fisher information is decreased with time. In the condition of $s<s_{c}$ which describes the Markovian environment, the quantum information always flows from the system to the environment. When $s>s_{c}$, the non-Markovianity of the reservoirs occurs because of the small positive values of $\frac {\partial F^{\mathrm{A}}_{\mathrm{max}}}{\partial t}>0$. For the temperature $\frac {\kappa_B T}{\hbar}=0.01$, the critical value of the bath parameter is about $s_{c}\approx 2.4$. In Fig. 1(b), the crossover between the Markovian dynamics and non-Markovian ones is clearly determined by the flow of the quantum Fisher information. If the bath parameter $s<s_{c}$, there is no dynamical crossover. For the non-Markovian environments of $s>s_{c}$, the values of $\frac {\partial F^{\mathrm{A}}_{\mathrm{max}}}{\partial t}>0$ will occur after a certain time $\omega_c\tau$. It means that the information can flow from the environment to the system after the times $t>\tau$. This non-Markovian dynamical behavior is represented by the region of $\textrm{NM}$. Before the time $\omega_c \tau$, the values of $F^{\mathrm{A}}_{\mathrm{max}}$ in the region of $\textrm{M}$ are monotonically decreased, which is referred to as the Markovian dynamics.

We want to know if the metrology strategy based on the quantum correlated states will have an advantage on improving the sensitivity of the phase estimation in the practical environment. The $N$-particle maximally entangled state is chosen to be the input state as
\begin{equation}
\label{eq:18}
\vert \Psi \rangle=\frac {1}{\sqrt{2}}(\Pi_{j=1}^{\otimes N}\vert 1 \rangle_{j}+\Pi_{j=1}^{\otimes N}\vert 0 \rangle_{j}).
\end{equation}
We assume that $N$ qubits are subject to the local environments independently. The total density matrix is obtained as
\begin{eqnarray}
\label{eq:19}
\rho^{N}(t)&=&\frac 12[\Pi_{j=1}^{\otimes N}\hat{\varepsilon}_{j}(\vert 1 \rangle_j \langle 1 \vert)+\Pi_{j=1}^{\otimes N}\hat{\varepsilon}_{j}(\vert 1 \rangle_j \langle 0 \vert) \nonumber \\
&+&\Pi_{j=1}^{\otimes N}\hat{\varepsilon}_{j}(\vert 0 \rangle_j \langle 1 \vert)+\Pi_{j=1}^{\otimes N}\hat{\varepsilon}_{j}(\vert 0 \rangle_j \langle 0 \vert)].
\end{eqnarray}
According tho the time-dependent density matrix written by Eq. (10), the dynamical map $\hat{\varepsilon}_{j}$ for the $\mathrm{ith}$ qubit can be expressed as
\begin{eqnarray}
\label{eq:20}
\hat{\varepsilon}_{j}(\vert 1 \rangle_j \langle 1 \vert)&=& a\vert 1 \rangle_j \langle 1 \vert+(1-a)\vert 0 \rangle_j \langle 0 \vert,\nonumber \\
\hat{\varepsilon}_{j}(\vert 1 \rangle_j \langle 0 \vert)&=& c\vert 1 \rangle_j \langle 0 \vert,\nonumber \\
\hat{\varepsilon}_{j}(\vert 0 \rangle_j \langle 1 \vert)&=& c^{\ast}\vert 0 \rangle_j \langle 1 \vert,\nonumber \\
\hat{\varepsilon}_{j}(\vert 0 \rangle_j \langle 0 \vert)&=& b\vert 1 \rangle_j \langle 1 \vert+(1-b)\vert 0 \rangle_j \langle 0 \vert.
\end{eqnarray}

Figure 2 shows the evolutions of the phase sensitivity described by the maximal mean values of the quantum Fisher information in non-Markovian depolarizing environment. When the number of the experiments $\nu=1$, the phase sensitivity can be given as $\Delta \phi\sim \frac {1}{\sqrt{\bar{F}_{\mathrm{max}}}}$ according to the quantum Cra\'{m}er-Rao theorem. It is clearly seen that the variances of the phase parameter based on quantum correlated states are always smaller than that gained by the product state. Here, we choose the product state as $\vert \Psi_{p}\rangle=\Pi_{j=1}^{\otimes N}\frac {1}{\sqrt{2}}(|1\rangle_j+|0\rangle_j)$. In the correlated-state metrology strategy, there exists a critical time $t_c$ where $\bar{F}_{\mathrm{max}}=1$. Before the critical time, we can obtain the ultimate sensitivity limit which is better than the standard quantum limit because of the high values $\bar{F}_{\mathrm{max}}>1$. However, the product-state metrology strategy cannot achieve the standard quantum limit owing to $\bar{F}_{\mathrm{max}}<1$. The maximal value of the variance arrives at the crossover time $t=\tau$. When $t>\tau$, the memory effects from the non-Markovian depolarizing environments can give rise to the decreasing of the phase sensitivity. After a long time, the sensitivity by using the quantum correlated state is nearly equivalent to that by using the product state.

The sudden change behavior in dynamics can be shown in the inset figure in Figure 2. Such behavior results from the competition of the two different kinds of the optimal $SU(2)$ rotations in either the $x-y$ plane or along $z$ direction. Previous to the sudden change time, the small variance of the phase parameter can be obtained by the $x-y$ plane rotation along $\vec{n}^{o}_{\parallel}$. While the optimal values of $(\Delta \phi)^2$ is achieved by the $z$ direction rotation along $\vec{n}^{o}_{\perp}$ at the time $t>t_s$. No sudden change behavior happens in the case of the product state because the value of $\bar{F}_{\mathrm{max}}$ is the same one with respect to the two kinds of optimal rotation directions.

On one hand, the occurrence of the sudden change behavior of the maximal quantum Fisher information is determined by the optimal selection of the rotation operations. On the other hand, $F_{\mathrm{max}}>1$ before the critical times is considered as a sufficient one for the existence of the entanglement \cite{Pezzeand09}. After the time $t>t_c$, the entanglement of the open quantum systems is nearly decreased to zero. Therefore, it is of interest to study the relationship between the critical times and sudden-changing ones. Trough the numerical calculation, we find that the vanishing of the entanglement is always more early than the happening of the sudden change behavior. The difference between the sudden-changing time and the critical time is plotted with the increasing of the number of qubits in Figure 3. It is found out that the scaling property can be obtained as $\omega_c(t_s-t_c)\propto \frac 1{N}$. When $N\rightarrow \infty$, the sudden-changing times are infinitely close to the critical times. For a very large number of entangled qubits, the sudden change behavior of the maximal quantum Fisher information can be used to characterize the existence of the entanglement. With respect to the maximal quantum Fisher information of large-$N$ entangled qubits, the selection of the optimal $SU(2)$ rotation along $\vec{n}^{o}_{\parallel}$ in the $x-y$ plane denotes the existence of the entanglement of open systems.

The impacts of the reservoirs on the sudden change dynamics of the phase sensitivity are shown in Figure 4. It is clearly seen that the sudden change behavior can occur in both Markovian ($s<s_c$) environments and non-Markovian ($s>s_c$) ones. For the cases of the smaller values of bath parameter $s$, the lower variance of phase estimation will be obtained. The values of the sudden-changing time $t_s$ are decreased with the increasing of the bath parameter. When $t<t_s$, the better resolution limit is achieved in the sub-Ohmic heat bath where the higher values of the maximal quantum Fisher information can be kept.

\section{Discussion}

We employ the spin-boson system to construct the practical non-Markovian depolarizing channel. The evolution of quantum Fisher information is applied to discriminate the Markovian dynamics and non-Markovian ones. There exists the temperature-dependent critical bath parameter $s_c$. Only if $s>s_c$, the crossover between the non-Markovian decoherence and Markovian one will exist. The sensitivity of the phase estimation was studied in both correlated-state metrology strategy and product-state metrology strategy. It is found out that the quantum correlated states can be used to improve the phase sensitivity. During the evolution, there are the critical time and sudden change time. Before the critical time, the resolution limit is intermediatly between the standard quantum limit (or shot-noise limit) and Heisenberg limit. The sudden change behavior will occur because of the competition of two different optimal rotations. The change of the reservoir parameter $s$ can lead to the variation of the sudden changing times. Moreover, for the large-$N$ cases, the sudden change times approximately equal to the critical times. For large-$N$ entangled states, we can obtain the maximal quantum Fisher information using the optimal $SU(2)$ rotation in the $x-y$ plane, which denotes the existence of the entanglement of open systems.

\section{ACKNOWLEDGEMENT}

This work is supported by the National Natural Science Foundation of China under Grant No. 11074184 and  No. 11174114. X. H. is financially supported from the China Postdoctoral Science Foundation funded project No. 2012M520494, the Basic Research Funds in Renmin University of China from the central government project No. 13XNLF03 and Research Project (No. 03040813) of Natural Science in Nantong University.

\newpage

{\Large \bf Figure Captions}

{\bf Fig. 1}

(a). The flow of the average of the maximal quantum Fisher information is plotted as functions of the bath parameter $s$ and scaled time $\omega_c t$ if $\frac {\kappa_B T}{\hbar}=0.01$ and $\omega_c=10\omega_0$; (b). The dynamical crossover related to the bath parameter $s$ is shown. The flow of the quantum Fisher information $\frac {\partial F^{\mathrm{A}}_{\mathrm{max}}}{\partial t}>0$ in the region $\textrm{NM}$ while $\frac {\partial F^{\mathrm{A}}_{\mathrm{max}}}{\partial t}<0$ in the region $\textrm{M}$.

{\bf Fig. 2}

The dynamics of the variance of the phase estimation $(\Delta \phi)^2$ is plotted in both correlated-state metrology strategy and product-state metrology strategy if $\frac {\kappa_B T}{\hbar}=0.01$, $\omega_c=10\omega_0$, $s=3$ and the number of qubits is $N=5$. The solid line represents the case of the maximally entangled state and dash-dot line denotes the case of the uncorrelated product state.

{\bf Fig. 3}

The scaling property of the critical time $t_c$ and sudden change time $t_s$ is plotted by the squares if $\frac {\kappa_B T}{\hbar}=0.01$, $\omega_c=10\omega_0$ and $s=3$. The dashed line represents the fit result.

{\bf Fig. 4}

For the strategy based on the maximally entangled state, the sudden change behavior can be shown with the change of the bath parameter $s$ if $\frac {\kappa_B T}{\hbar}=0.01$, $\omega_c=10\omega_0$ and $N=5$.

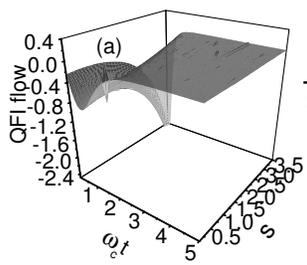
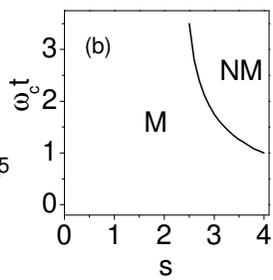
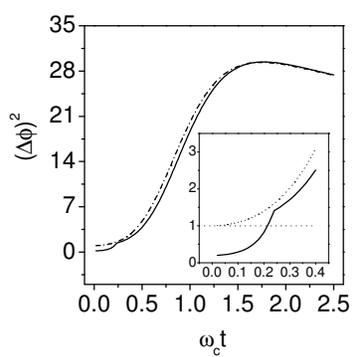
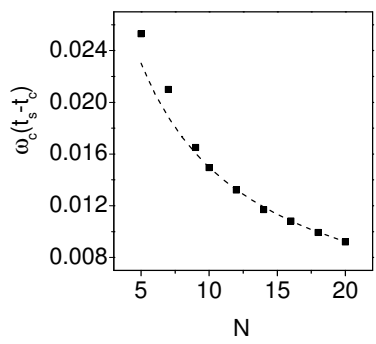
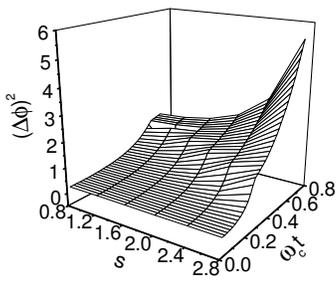
\end{document}